\documentclass[twocolumn,prl,aps,floats,amsmath,10pt,amssymb,superscriptaddress, footinbib, showpacs]{revtex4-1}
\usepackage{epsfig}
\usepackage{graphicx}
\usepackage{hyperref}
\usepackage[usenames,dvipsnames]{xcolor}
\usepackage[final]{pdfpages}

\newcommand{\PRLsec}[1]{\noindent\textit{#1}---}

\newcommand*{\rom}[1]{\expandafter\@slowromancap\romannumeral #1@}

\begin{document}
\title{Nonlinear Oscillations and Bifurcations in Silicon Photonic Microresonators}
\date{\today}

\author{Daniel M. Abrams}
\email{dmabrams@northwestern.edu}
\affiliation{Department of Engineering Sciences and Applied Mathematics, Northwestern University, Evanston, IL 60208}
\affiliation{Northwestern Institute on Complex Systems, Northwestern University, Evanston, IL 60208}
\affiliation{Department of Physics \& Astronomy, Northwestern University, Evanston, IL 60208}

\author{Alex Slawik}
\email{AlexanderSlawik2015@u.northwestern.edu}
\affiliation{Department of Engineering Sciences and Applied Mathematics, Northwestern University, Evanston, IL 60208}

\author{Kartik Srinivasan}
\email{kartik.srinivasan@nist.gov}
\affiliation{Center for Nanoscale Science and Technology, National Institute of Standards and Technology, Gaithersburg, MD 20899}


\begin{abstract}
Silicon microdisks are optical resonators that can exhibit surprising nonlinear behavior. We present a new analysis of the dynamics of these resonators, elucidating the mathematical origin of spontaneous oscillations and deriving predictions for observed phenomena such as a frequency comb spectrum with MHz-scale repetition rate. We test predictions through laboratory experiment and numerical simulation.
\end{abstract}

\maketitle

A remarkable self-oscillation \cite{ref:orozco1984, ref:segard1988, ref:joshi2003, ref:armaroli2011, ref:gu2012, ref:kwon2013} effect has recently been observed in silicon photonic microresonators \cite{ref:Priem_Baets_pulsation, ref:johnson2006, ref:soltani2012}, where excitation of the device with a continuous wave input field can yield a periodically time-varying output field. Here, we present a new analysis of bifurcations and oscillations in silicon microresonators, predicting the location and period of oscillation in parameter space.

Previous work has examined this phenomenon through direct numerical integration \cite{ref:johnson2006} and two-timescale approximation~\cite{ref:soltani2012}.  By analyzing the structure of the coupled equations and the timescales over which different physical effects occur, we are able to reduce the dimensionality of the system and derive approximate closed-form expressions for characteristic physical phenomena. As one example, our analysis predicts that the intracavity field can exhibit a stable limit cycle manifested by a comb of equally spaced frequency components. 

The physical insight derived from this approach may be valuable in efforts to make use of these devices as compact, optically-driven oscillators.  More generally, improved understanding of nonlinear phenomena in silicon resonators is important given their wide range of applications in
photonics~\cite{ref:Xu_Lipson_modulator,*ref:Green_Vlasov_modulator,*ref:Reed_modulator_review,ref:Xia_Vlasov_Si_buffer,*ref:Melloni_delay_line,*ref:Mookherjea_CROW,ref:Foster_Gaeta_gain,*ref:Foster_Lipson_FWM_microring, ref:mabuchi2009}.

\begin{figure}[th]
  \includegraphics[width=8.6cm, height=4.66cm]{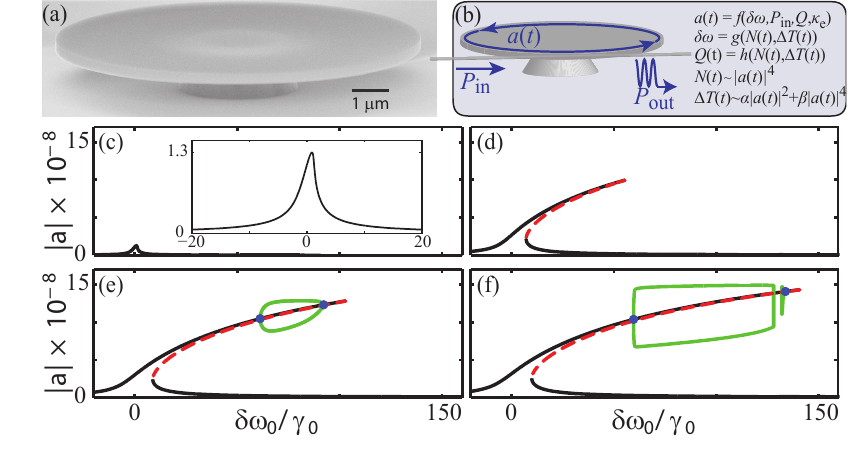}
  \caption{(a) Scanning electron microscope image of a silicon microdisk resonator. (b) Schematic of the physical system, in which a continuous-wave input field results in a periodically oscillating output field. (c)--(f) Steady-state resonance curves for the intracavity field amplitude as a function of normalized detuning.  (c) At low power the curve is stable and single valued. (d) As power increases,  nonlinear effects grow and the resonance curve bends over, leading to an unstable middle branch (dashed red). (e) Further increase in pump power leads to two simultaneous Hopf bifurcations (blue dots) and the birth of a stable limit cycle (envelope shown in green). (f) When the limit cycle grows sufficiently large, it collides with the middle branch and is destroyed via a homoclinic bifurcation (this collision occurs in four dimensions, and is not visible in this projection.) Pump powers: 0.71~$\mu$W, 45~$\mu$W, 86~$\mu$W, and 120~$\mu$W.}
  \label{fig:rescurves}
\end{figure}

\PRLsec{Physical system and model}%
The physical system we study is a microdisk cavity (Fig.~\ref{fig:rescurves}(a)) coupled to a single mode optical waveguide. The waveguide is driven with a continuous-wave laser at a specified frequency detuning with respect to a microdisk optical mode. For simplicity, we neglect backscattering effects common in these types of resonators~\cite{ref:Weiss_backscattering}, and assume that the forward propagating mode of the waveguide excites only the clockwise traveling-wave mode of the microdisk \cite{ref:harayama1999}, which in turn couples back out to the forward propagating mode.  
Our analysis neglects the Kerr nonlinearity, which has been the focus of considerable experimental~\cite{delhaye2007, *foster2011, *matsko2013} and theoretical work~\cite{lugiato1987, *leo2010, *chembo2010, *matsko2011, *herr2013} in the context of parametric oscillation and frequency comb generation.  It also neglects Raman scattering, and instead, focuses on the role of two-photon absorption (TPA).
As summarized in Supplemental Material and in Fig.~\ref{fig:rescurves}(b), a strong enough intracavity field produces two-photon absorption in the silicon material, resulting in heating and thermo-optic dispersion, as well as the generation of free carriers, which cause additional absorption (FCA) and dispersion.  The change in the optical loss rate and laser-cavity detuning caused by these effects means that the intracavity field $a(t)$ is coupled to the cavity temperature change $\Delta T(t)$ and the free carrier population $N(t)$.

The physical effects summarized above are described by the following set of coupled differential equations \cite{ref:johnson2006}:
\begin{subequations}
  \begin{widetext}
    \label{eq:base}
    \begin{align}
    \frac{da}{dt} =& -\frac{1}{2} \left( \gamma_{0} + \gamma_{e} + \frac{\Gamma_{TPA}\beta_{Si}c^{2}}{V_{TPA}n_{g}^{2}} \left|a(t)\right|^{2} + \frac{\sigma_{Si}c N(t)}{n_{g}} \right)a(t)+ i\left(\frac{\omega_{0}\frac{dn_{Si}}{dT}\Delta T(t)}{n_{Si}} + \frac{\omega_{0} \frac{dn_{Si}}{dN}N(t)}{n_{Si}} -\delta \omega_{0}\right) a(t) - i \kappa P_{in}^{1/2} ~, \label{eq:base1}
    \end{align}
  \end{widetext}
  \begin{align}
    \frac{dN}{dt} =& -\gamma_{fc}N(t)+\frac{\Gamma_{FCA}\beta_{Si}c^{2}}{2 \hbar \omega_{0}n_{g}^{2}V_{FCA}^{2}}\left|a(t)\right|^{4}~, \\
    \frac{d \Delta T}{dt} =& -\gamma_{Th}\Delta T(t)+ \frac{\Gamma_{disk}}{\rho_{Si}c_{p}V_{disk}}\bigg(\gamma_{lin} \nonumber \\
    +& \frac{\sigma_{Si}cN(t)}{n_{g}}+\frac{\Gamma_{TPA}\beta_{Si}c^{2}}{V_{TPA}n_{g}^{2}}\left|a(t)\right|^{2} \bigg)\left|a(t)\right|^{2} ~,
  \end{align}
\end{subequations}
(see Supplemental Material Section S2).

Key parameters that we allow to vary include the input laser's detuning frequency $\delta\omega_{0}=\omega_{0}-\omega_{in}$ (sign is opposite of typical optics convention) and the input power $P_{in}$. We refer the reader to Supplemental Material sections S2 and S3 for details on the system and the values of parameters. For simplicity in analysis, we separate Eq.~\eqref{eq:base1} into real and imaginary parts, then nondimensionalize to obtain
\begin{subequations}
  \label{eq:4D}
  \begin{align}
  \frac{dU}{d\tau} =-&A_{1}U - A_{2} S^2 U(U^{2}+V^{2}) - A_{3}\eta U + A_{4} \eta V \nonumber \\
    +& A_{5} x V-A_{6} \theta V~, \label{eq:4Da} \\
  \frac{dV}{d\tau} =-&A_{1}V - A_{2} S^2 V(U^{2}+V^{2}) - A_{3}\eta V - A_{4} \eta U \nonumber \\
    -& A_{5} x U+A_{6} \theta U - A_{7}~, \label{eq:4Db} \\
  \frac{d\eta}{d\tau} =-&A_{8}\eta + A_{9} S^4 (U^{2}+V^{2})^{2}~,  \label{eq:4Dc} \\
  \frac{d\theta}{d\tau} =-&A_{10}\theta + A_{11} S^2 (U^{2}+V^{2}) + A_{12}S^4(U^{2}+V^{2})^{2} \nonumber \\
    +& A_{13}S^2 \eta(U^{2}+V^{2})~, \label{eq:4Dd}
  \end{align}
\end{subequations}
where $\tau = \frac{\gamma_{0}}{\sqrt{Q}}t$, $U = \frac{\sqrt{\omega_{0}}}{6 Q^{1/4}\sqrt{P_{in}}} \text{Re}(a)$, $V = \frac{\sqrt{\omega_{0}}}{6 Q^{1/4} \sqrt{P_{in}}} \text{Im}(a)$, $\eta = \frac{V_{eff}}{Q} N$, and $\theta=\frac{c_p}{\gamma_0^2 \sigma_{Si} Q} \Delta T$ are dimensionless real variables of order 1, $A_1$ through $A_{13}$ are positive real constants (see Supplemental Material), and $x=\delta\omega_{0}/\gamma_0$, $S=(\beta_{Si}\omega_0 c^{-1}Q^2 P_{in})^{1/2}$ are the nondimensional corollaries to control parameters $\delta\omega_{0}$ and $P_{in}$.

\PRLsec{Regions of Oscillation and Bistability}%
Figure \ref{fig:rescurves} shows the field amplitude $\left|a\right|$ vs detuning for various driving powers. As power increases, the resonance curve becomes multivalued---bistability and hysteresis becomes possible. At a critical pump power, two simultaneous Hopf bifurcations occur (two pairs of eigenvalues cross the imaginary axis), destabilizing part of the upper branch and leading to the birth of a limit cycle between the two Hopf bifurcations.  As pump power is further increased, this limit cycle collides with the unstable fixed point (middle branch), undergoing a homoclinic bifurcation that destroys its stability within a range of detunings---see Fig.~\ref{fig:rescurves} panel (f).
\begin{figure}[th]
  \includegraphics[width=8.6cm, height=5.033cm]{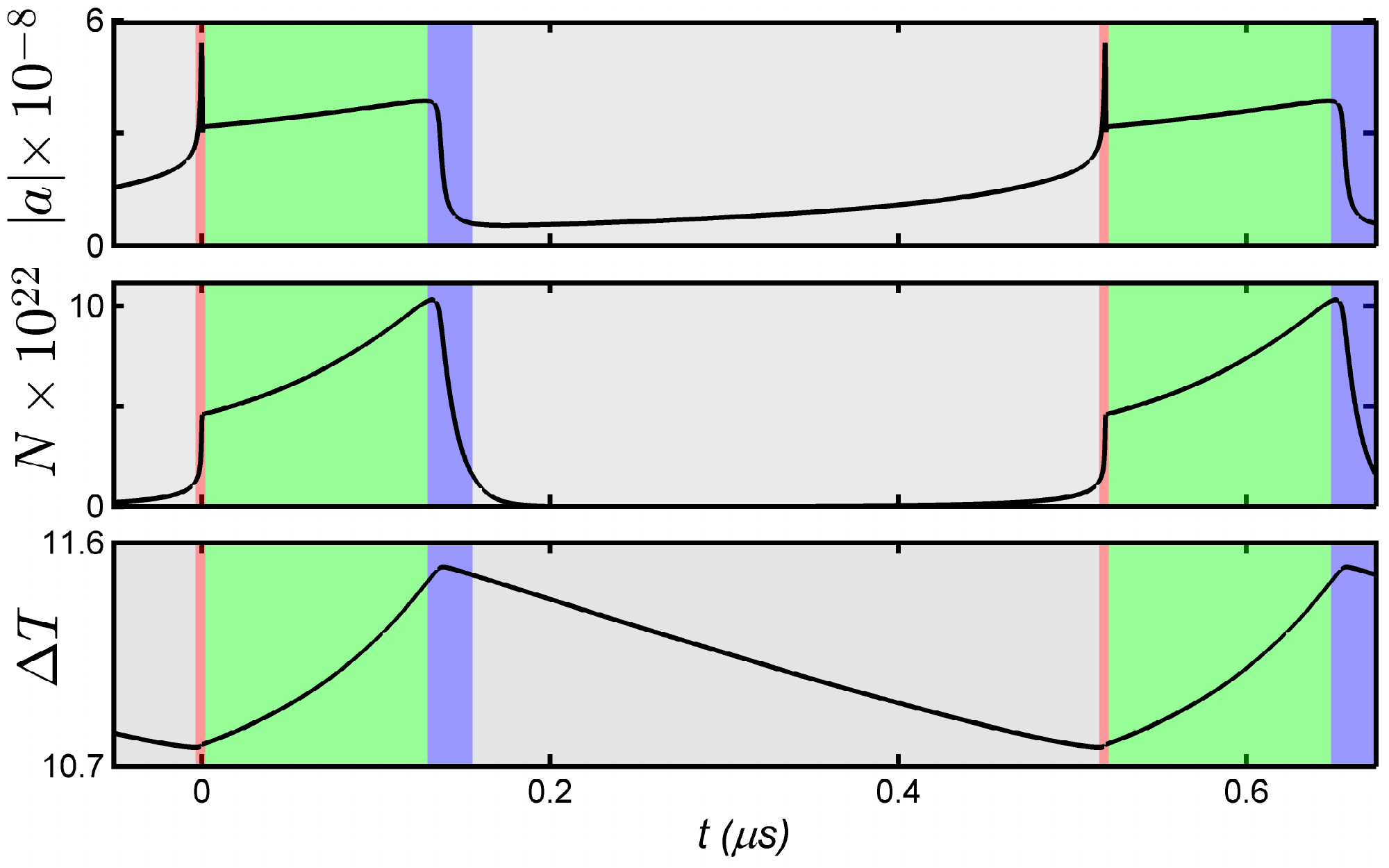}
  \caption{Limit cycle oscillation. Panels show (a) field amplitude $\left|a\right|$, (b) free carrier population $N$ and (c) temperature change $\Delta T$ vs. time. Colors indicate different stages of limit cycle. Pump power is 1~mW ($S = 56$) and pump detuning is 0.84 nm above resonance ($\delta\omega_{0}/\gamma_{0}=168$).}
  \label{fig:4D_lc}
\end{figure}

\PRLsec{Time Domain Behavior}%
Figure \ref{fig:4D_lc} shows the periodic behavior of the system with high pump power and a stable limit cycle. It consists roughly of four stages and can be interpreted physically as follows: the first stage (red) starts at minimum temperature and is driven by rapid TPA. A sharp spike in the field is tempered by linear and nonlinear optical losses and the blue shift of the disk's resonant frequency due to a denser free carrier population. The free carrier population stabilizes when free carrier recombination ($\gamma_{fc}$) balances with free carrier generation via TPA. Thermal decay ($\gamma_{Th}$) happens more slowly, so cavity temperature doesn't equilibrate during the spike. The second stage (green) is driven by an increasing temperature red-shifting the disk's resonant frequency, and consists of steady increases in all variables. A critical temperature is reached (blue), and both the field and free carrier population collapse in conjunction with a sudden drop in TPA. The fourth stage (gray) takes up most of the limit cycle and consists of low activity in the disk while the temperature decreases smoothly.

\begin{figure}[th]
  \includegraphics[width=8.6cm, height=3.44cm]{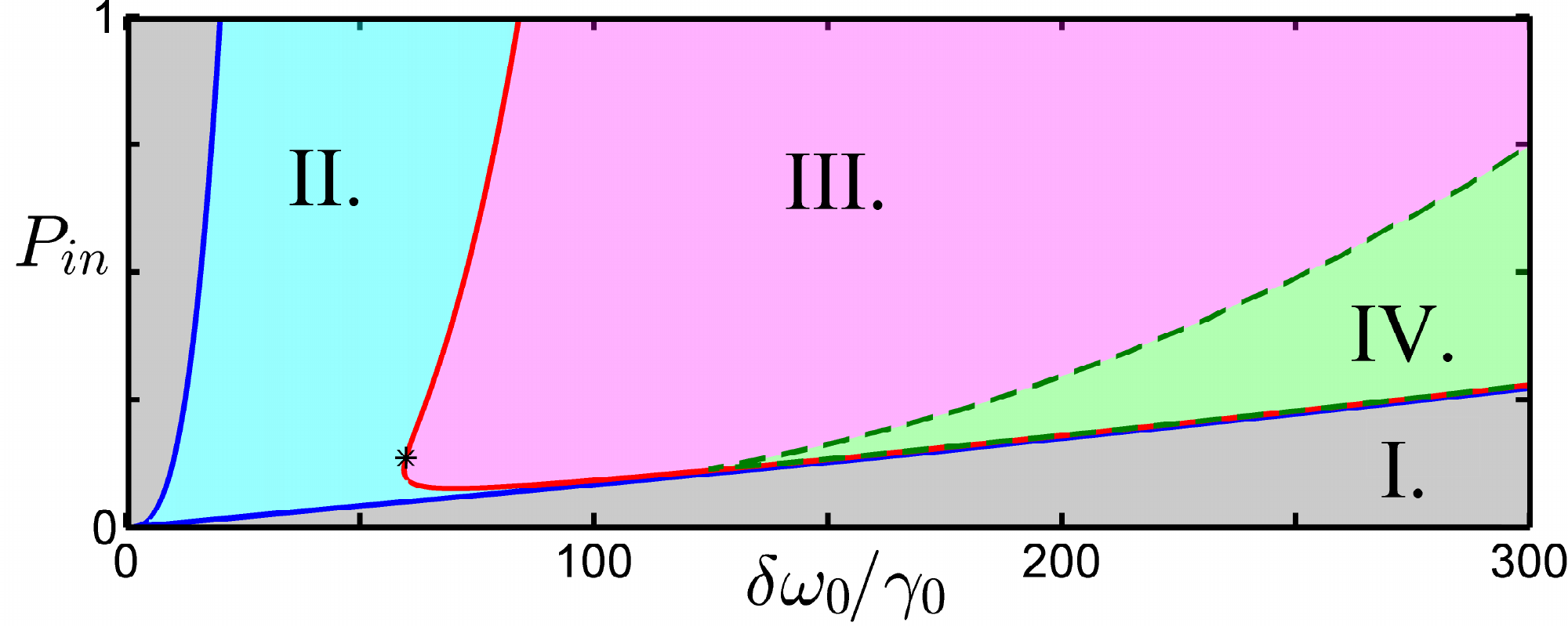}
  \caption{Phase space diagram of system in parameter space of power ($P_{in}$) and detuning ($\delta\omega_{0}/\gamma_{0}$). Region I (gray): monostable (one stable equilibrium), region II (blue): bistable (two stable, one unstable equilibria), region III (pink): stable oscillations (one stable, two unstable equilibria), region IV (green): monostable (one stable, two unstable equilibria).  Blue boundary: saddle-node bifurcation, red boundary: Hopf bifurcation, dashed green boundary: homoclinic bifurcation. Black asterisk indicates point where Hopf bifurcation goes from subcritical to supercritical. Power ranges from 0~mW to 1~mW, detuning ranges from 0~nm to 1.5~nm above resonance.}
  \label{wedge}
\end{figure}

Figure \ref{wedge} shows bifurcations that occur in the parameter space of $\delta\omega_{0}/\gamma_{0}$ and $P_{in}$. The limit cycle is ``born'' in parameter space on the boundary defined by the Hopf-condition (red line) with non-zero period $T$. At powers above a threshold (black asterisk), the limit cycle transitions from supercritical (born with zero amplitude) to sub-critical (born with finite amplitude). In the low power limit, we use a local asymptotic expansion about the Hopf condition to accurately approximate the limit cycle (see Supplemental Material). At higher power we use a multiple-time-scale analysis to ultimately reduce the limit cycle to a one-dimensional relaxation oscillation, and predict the Hopf and homoclinic bifurcations \cite{ref:kuznetsov1998elements}.

\PRLsec{Multiple Time Scales}%
In their analysis, Johnson \textit{et al}.~suggested that the observed limit cycle can be separated into fast and slow time scales~\cite{ref:johnson2006}. Soltani \textit{et al}.~carried out a two-time-scale approximation by assuming changes in temperature are much slower than changes in other variables~\cite{ref:soltani2012}; these time scales are apparent in Fig.~\ref{fig:4D_lc}. Here we extend that idea to a convenient approximation in terms of three well-separated time scales. Specifically, the equations governing the field \eqref{eq:4Da}-\eqref{eq:4Db}, the free carriers \eqref{eq:4Dc}, and the temperature \eqref{eq:4Dd} each appear to operate on a different time scale. 

Our approach is based upon order of magnitude comparison between the model's coefficients (see Supplemental Material). The ratio of the coefficients in Eqs.~\eqref{eq:4Da} and \eqref{eq:4Db} to $A_{1}$ is at least of order one, while the ratio of the coefficients in Eqs.~\eqref{eq:4Dc} and \eqref{eq:4Dd} to $A_{1}$ is much less than one \cite{Note1}, as long as $A_{3}\gg A_{8}$, which implies that $ \sigma_{Si}c Q\gg V_{eff}n_{Si}\gamma_{fc}$ (a less restrictive but necessary relation is $A_{1}\gg A_{8}$, or $\gamma_{0}\gg \gamma_{fc}$). When these relations hold, Eqs.~\eqref{eq:4Da} and \eqref{eq:4Db}, Eq.~\eqref{eq:4Dc}, and Eq.~\eqref{eq:4Dd} evolve on time scales $\tau_{1}=\gamma_{0} t$, $\tau_{2}=\gamma_{fc} t$, and $\tau_{3}=\gamma_{Th} t$ respectively, with $\tau_{1}\gg\tau_{2}$ and $\tau_{1}\gg\tau_{3}$.

Taking the free carrier population $\eta$ and the temperature change $\theta$ to be constant, the solution to equations \eqref{eq:4Da} and \eqref{eq:4Db} approach fixed points $U^{\star}=\frac{c_2 A_7}{c_1^2+c_2^2}$, $V^{\star}=\frac{c_1 A_7}{c_1^2+c_2^2}$ exponentially fast, where $c_{1}=A_1+A_3 \eta$ and $c_{2}=-A_5 x +A_6 \theta -A_4 \eta$. Numerical simulation verifies that the values of $U$ and $V$ are well approximated by these fixed points during the limit cycle. We conclude that the apparent fast dynamics observed in Fig.~\ref{fig:4D_lc} are slaved to the dynamics of the free carrier population.

\begin{center}
  \begin{figure}[th]
  \includegraphics[width=8.6cm, height=5.375cm]{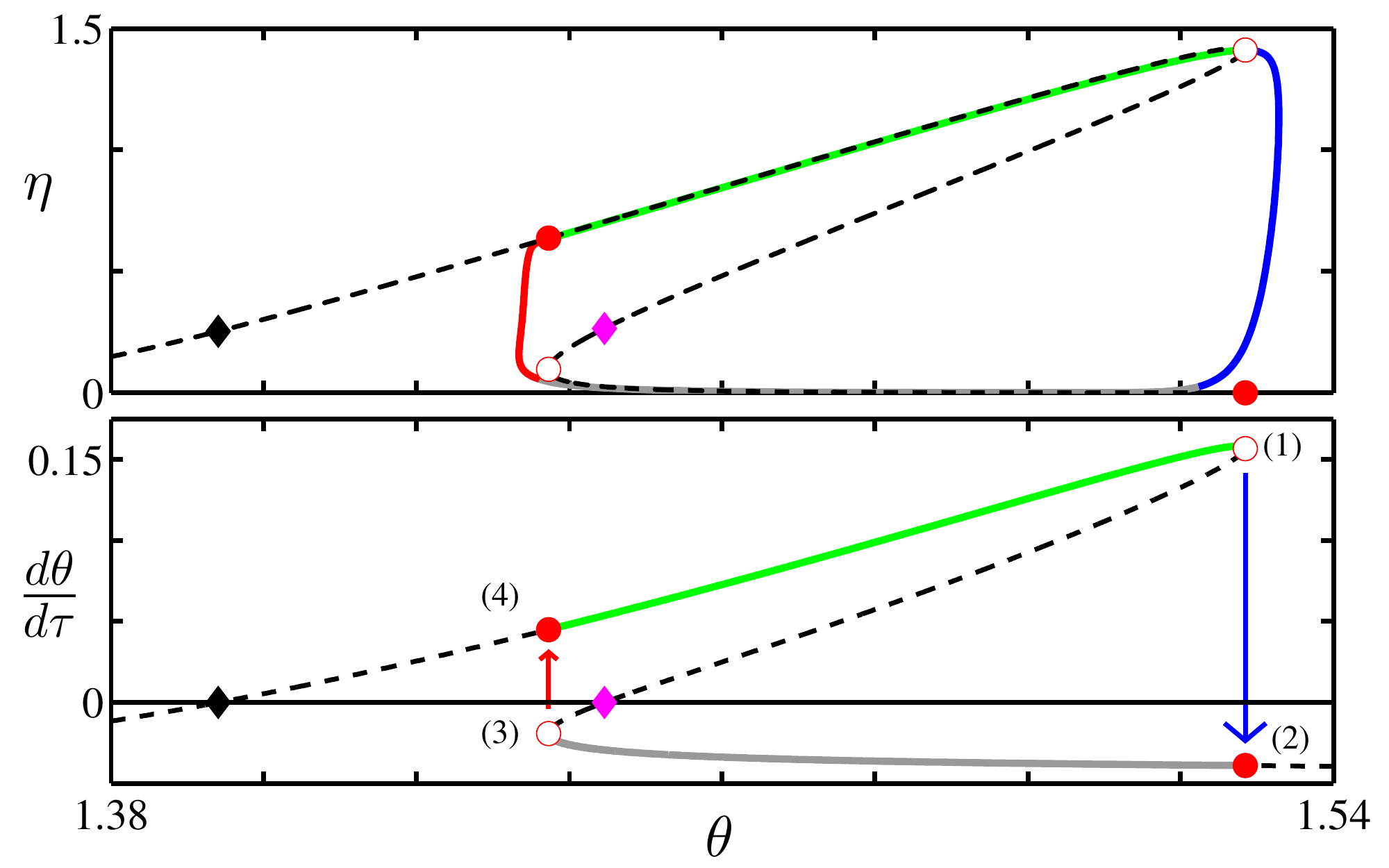}
  \caption{Limit cycle for 2D and 1D reduction in the space of nondimensional free carrier population $\eta$ and temperature change $\theta$ (Eqns.~\eqref{eq:2D} and \eqref{eq:thsoln}/\eqref{eq:thdotsoln} respectively). Top panel: stable periodic solution to 2D model (solid), nullcline $d\eta/d\tau=0$ (dashed), unstable fixed points (filled diamonds), and points of interest in the 1D reduction (filled and open circles). Lower panel: same labeling scheme, solid lines represent branches of nullcline corresponding to 1D limit cycle, arrows indicate instantaneous jumps. Input power is 1~mW ($S = 56$) and pump detuning is 0.84~nm above resonance ($x=168$). Color coding indicates portion of cycle with same scheme as Fig.~\ref{fig:4D_lc}. }
  \label{fig:1D_2D}
  \end{figure}
\end{center}

Thus, assuming field variables $U$ and $V$ reach equilibrium nearly instantaneously in response to changes in $\eta$ and $\theta$, system \eqref{eq:4D} reduces to
\begin{subequations}
\label{eq:2D}
\begin{align}
  \frac{d\eta}{d\tau} =& -A_8 \eta + {\scriptstyle \frac{A_9 A_{7}^4 S^4}{\left[ (-A_5 x + A_6 \theta-A_4 \eta)^{2}+(A_1+A_3 \eta)^{2} \right]^2}} \label{slo_eq1} \\
  \frac{d\theta}{d\tau} =& -A_{10} \theta + {\scriptstyle \frac{(A_{11}A_{7}^2+A_{13}A_{7}^2 \eta)S^2}{(-A_5 x + A_6 \theta-A_4 \eta)^{2}+(A_1+A_3 \eta)^{2}}} \nonumber \\
  +& \frac{A_{12}A_{7}^4 S^4}{ \left[ (-A_5 x + A_6 \theta-A_4 \eta)^{2}+(A_1+A_3 \eta)^{2} \right]^2}~. \label{slo_eq2}
\end{align}
\end{subequations}
As expected, this 2D system behaves nearly identically to the 4D system when the above assumptions are satisfied. Figure \ref{fig:1D_2D} shows the limit cycle in the phase plane of $\eta$ and $\theta$ along with the nullcline $\frac{d \eta}{d \tau}=0$ (dashed).

Note that in Fig.~\ref{fig:1D_2D} the value of $\eta$ is nearly always either on the nullcline, or changing rapidly with respect to $\theta$. That is, $\frac{d\eta}{d\tau}\gg\frac{d\theta}{d\tau}$ when not on a nullcline. This observation allows us to simplify the system further through a second separation of time scales: we'll assume that $\eta$ is nearly always at a fixed point. This is valid when $A_{10} \ll A_{8}$ and $A_{10}\ll A_{9} S^4$, with the former relation implying that $\gamma_{fc} \gg \gamma_{Th}$. For a disk resting on a pedestal of SiO$_{2}$, $\gamma_{Th} \approx 0.2\ \textrm{MHz}$, while $\gamma_{fc}$ is typically $\mathcal{O}(100\ \textrm{MHz})$ \cite{ref:johnson2006}, so the assumption should be valid in our experiments. As long as these conditions and the fast field conditions hold, Eqs.~\eqref{eq:4Da} and \eqref{eq:4Db}, Eq.~\eqref{eq:4Dc}, and Eq.~\eqref{eq:4Dd} evolve on time scales $\tau_{1}=\gamma_{0} t$, $\tau_{2}=\gamma_{fc} t$, and $\tau_{3}=\gamma_{Th} t$ respectively, with $\tau_{1}\gg\tau_{2}\gg\tau_{3}$ ($\gamma_{0}\gg\gamma_{fc}\gg\gamma_{Th}$). 

Setting Eq.~\eqref{slo_eq1} equal to zero gives the following parameterization in terms of $\eta$:
\begin{eqnarray}
  \label{eq:thsoln}
  \theta  &=& \frac{A_{5} x+A_{4} \eta}{A_{6}} \pm \frac{\sqrt{f(\eta)}}{A_{6}\eta}~, \\
  f(\eta) &=& - A_{3}^2 \eta^4-2 A_{1} A_{3} \eta^{3}-A_{1}^2\eta^2 +S^2\sqrt{\frac{A_{7}^4 A_{9}}{A_{8}}}\eta^{3/2}~, \nonumber
\end{eqnarray}
and plugging Eq.~\eqref{eq:thsoln} into Eq.~\eqref{slo_eq2} gives $\dot{\theta}=\frac{d\theta}{d\tau}$ in terms of $\eta$:
\begin{eqnarray}
  \label{eq:thdotsoln}
  \dot{\theta} =&& -\frac{A_{10} A_5 x}{A_6}+\frac{\sqrt{A_{8}}A_{11}}{\sqrt{A_9}}\sqrt{\eta}+\frac{\sqrt{A_{8}} A_{13}}{\sqrt{A_9}}\eta^{3/2} \\
  && + \left(\frac{A_8 A_{12}}{A_9}-\frac{A_{10}A_{4}}{A_6}\right)\eta  \pm  \frac{A_{10} \sqrt{f(\eta)}}{A_{6} \eta} \nonumber
\end{eqnarray}
Figure \ref{fig:1D_2D} illustrates this 1D reduction of the 2D limit cycle. The limit cycle occurs in the region of the graph that is multivalued. The boundaries of the region (maximum and minimum values of $\theta$) mark transition points between the two solution curves.

\PRLsec{Estimating the Period of the Limit Cycle}%
The 1D reduction assumes that the transition between ``jump'' and ``collection'' points is instantaneous, separating the limit cycle into four sections: two fast (red and blue sections of Figs.~\ref{fig:4D_lc} and \ref{fig:1D_2D}) and two slow (gray and green sections of Figs.~\ref{fig:4D_lc} and \ref{fig:1D_2D}). Integrating $\frac{1}{\dot{\theta}(\eta)} \frac{d \theta}{d \eta}$ with respect to $\eta$ along the nullcline from the collection points to the jump points gives the period of the 1D limit cycle. Using approximations to the phase plane branches, we found
\begin{eqnarray}
  T &\approx& 2 \frac{\theta^{*}_{1} - \theta^{*}_{3}}{(\dot{\theta}^{*}_{2} - \dot{\theta}^{*}_{3})^{2}} \left[ \dot{\theta}^{*}_{3} \ln{ \left(\frac{ \dot{\theta}^{*}_{3}}{\dot{\theta}^{*}_{2}} \right) } + (\dot{\theta}^{*}_{2}-\dot{\theta}^{*}_{3}) \right] \nonumber \\
  &+& \frac{\theta^{*}_{1}- \theta^{*}_{3}}{\dot{\theta}^{*}_{1} - \dot{\theta}^{*}_{4}} \ln{ \left( \frac{\dot{\theta}^{*}_{1}}{\dot{\theta}^{*}_{4}} \right) }~,
  \label{eq:Ttheory}
\end{eqnarray}
where starred variables indicate known jump and collection points (subscripts refer to numbered critical points in Fig.~\ref{fig:1D_2D}).

\PRLsec{Limits of Oscillation}%
The 1D reduction yields intuitive and simple expressions for the limits of oscillation with respect to detuning. Equations \eqref{eq:thsoln} and \eqref{eq:thdotsoln} imply that changes in detuning simply translate the limit cycle. With increasing detuning, the onset of oscillations occurs when the bottom left ``elbow'' Fig.~\ref{fig:1D_2D} (open circle) crosses the $\theta$ axis ($\dot{\theta}^{*}_{3}=0$). The collapse of oscillations through homoclinic bifurcation occurs when the limit cycle collides with the nearby unstable fixed point ($\dot{\theta}^{*}_{4}=0$). The period of the limit cycle diverges near this instability. By using Eqs.~\eqref{eq:thsoln} and \eqref{eq:thdotsoln} we can express the bounds of oscillation in terms of all free parameters.

\begin{figure}[ht]
  \includegraphics[width=8.6cm, height=3.91cm]{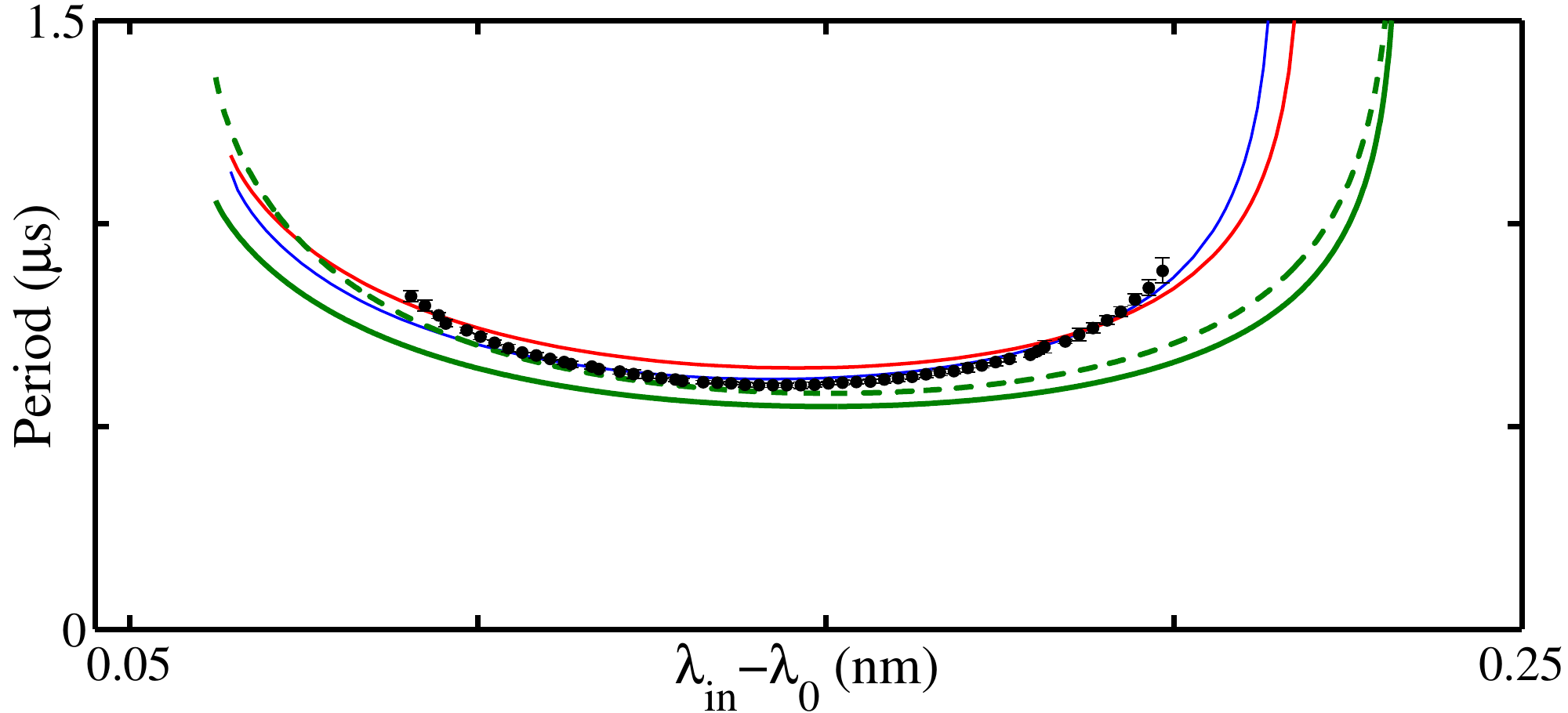}
  \caption{Existence and period of limit cycle. Comparison of 4D (blue), 2D (red), and 1D (green) models from equations \eqref{eq:4D}, \eqref{eq:2D}, \eqref{eq:thsoln}/\eqref{eq:thdotsoln}, and \eqref{eq:Ttheory} to experiment (black points; vertical error bars stem from Lorentzian fits to determine frequency peak locations, and represent one standard deviation in the comb spacing)
  for $P_{in}=400\ \mu\textrm{W}$, $\lambda_{0}=1609$~nm, $Q=6\times10^{5}$, $V_{eff}$ = 60 $\left(\frac{\lambda_{0}}{n_{Si}}\right)^{3}$ (fit), $\gamma_{lin}/\gamma_{0}=0.53$ (fit), $\gamma_{Th}=1.4\times10^{5}$Hz (fit), and $\gamma_{e}/\gamma_{0}=0.08$. Solid line=numerical solution, dashed=analytical approximation.}
  \label{fig:Period_comp}
\end{figure}

Figure \ref{fig:Period_comp} compares the predictions of the 4D model (Eq.~\eqref{eq:4D}), the 2D model (Eq.~\eqref{eq:2D}), and the 1D model (Eqs.~\eqref{eq:thsoln}/\eqref{eq:thdotsoln} and \eqref{eq:Ttheory}) to laboratory data (see Supplemental Material), indicating that they capture the dependence of the period of oscillation on detuning. The 1D reduction overestimates the detuning at which the homoclinic bifurcation occurs due to failure to capture the ``overshoot'' near instantaneous jumps between branches.

The dependency of period on other system parameters is generally similar to Fig.~\ref{fig:Period_comp}.  Increasing the strength of nonlinear terms usually increases the period of oscillation.  In general, changes in period are more severe at the bounds of the limit cycle in parameter space (red and dashed green curves in Fig.~\ref{wedge}, see Supplemental Material section S6 for a numerical survey).

\begin{figure}[th]
  \includegraphics[width=8.6cm, height=3.44cm]{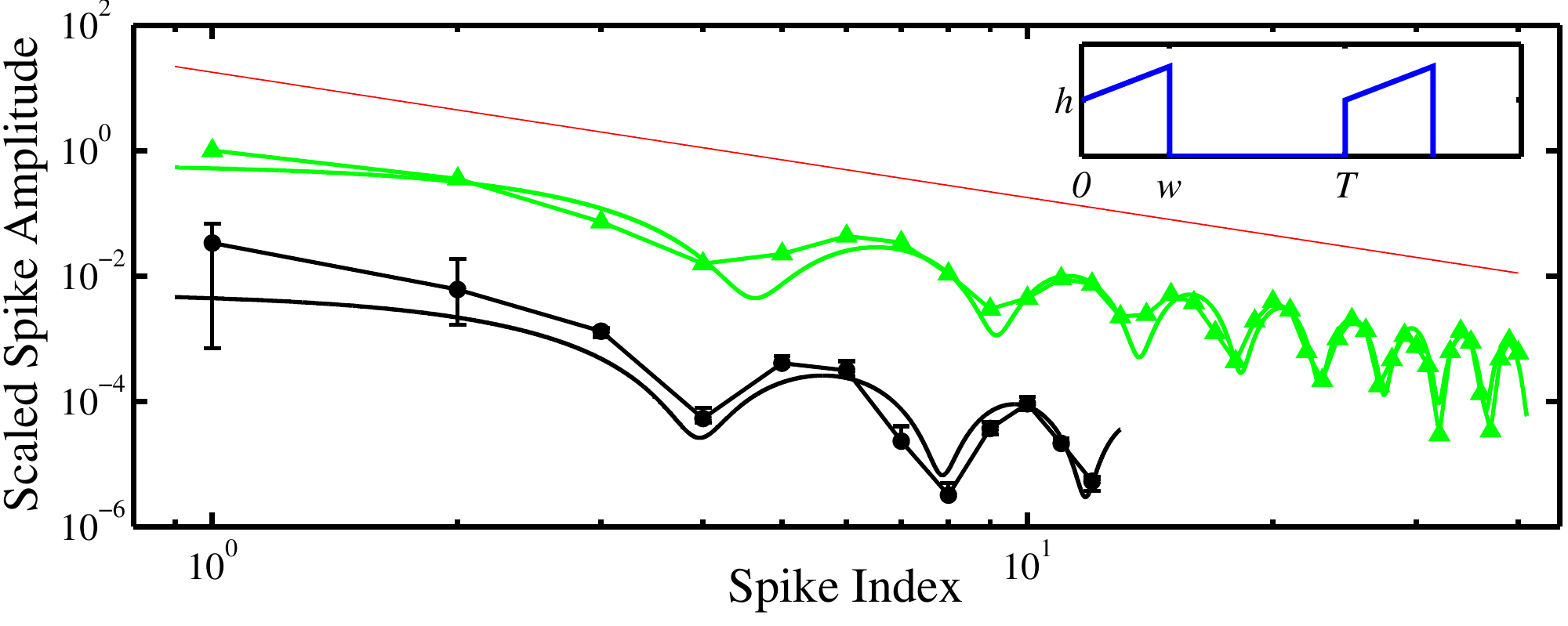}
  \caption{Frequency comb envelope. The frequency comb's decay for both experimental data (connected black circles) and 4D numerics (connected green triangles) is compared with a best fit saw-tooth pulse wave (solid lines, see inset). Primary spike height is arbitrarily scaled for visual purposes. Red reference line shows $n^{-2}$. Error bars indicate 90$\%$ bootstrap confidence interval derived from Lorentzian fit.}
  \label{freq_comb_fit}
\end{figure}

\PRLsec{Frequency Comb}%
The self-sustained oscillations of the field inside the cavity produce a frequency comb with spacing on the order of 1 MHz~\cite{ref:johnson2006} (data presented in Supplemental Material). This spacing corresponds to the frequency of the limit cycle, and multiple lines appear since multiple Fourier modes are necessary to represent its non-sinusoidal shape. The amplitude of successive peaks in the comb can be deduced from the structure of the time-domain oscillation.  The spike and subsequent abrupt slope-change visible in Fig.~\ref{fig:4D_lc} is primarily responsible for generating the higher harmonics in the comb and suggest the use of a modified pulse wave for approximate theoretical prediction of the comb envelope. We find that the frequency comb's higher harmonics decay according to a power law with $n^{-x}$ where $n$ is the index of the harmonic and $x \approx 2$. The power spectrum of a sawtooth pulse wave oscillates about the decay rate of $n^{-2}$ with an oscillatory period (in spikes) of $T/w$, where $T$ is the fundamental period and $w$ is the pulse width. Figure \ref{freq_comb_fit} shows the fit of both the data and the 4D numerics to the frequency comb of a sawtooth pulse wave.

\PRLsec{Discussion of Results}%
We have presented a new approach to modeling the multi-scale oscillatory behavior brought on by nonlinear absorption and dispersion in silicon microdisks. Perturbation theory allows us to reduce dimensionality and gain insight into the underlying dynamics of this nonlinear system, even producing analytic predictions for key properties of the system and key transitions and behavior. The heart of the analysis lies in the separation of time scales between optical, electro-optical, and thermal effects which are characteristic of multiple optoelectronic devices, including the silicon microdisks considered in our work.

\PRLsec{Acknowledgements}%
The authors thanks V.~Akysuk, L.~Chen, and H.~Miao for helpful discussions regarding fabrication of silicon microdisk devices.


\pagestyle{empty} \hspace{0mm} \newpage \hspace{0mm} \newpage


\section*{Supplementary material}

# Nonlinear Oscillations and Bifurcations in Silicon Microdisk Resonators: Supplemental Material

Daniel M. Abrams,[1, 2, 3, *] Alex Slawik,[1, †] and Kartik Srinivasan[4, ‡]

[1]Department of Engineering Sciences and Applied Mathematics, Northwestern University, Evanston, IL 60208
[2]Northwestern Institute on Complex Systems, Northwestern University, Evanston, IL 60208
[3]Department of Physics & Astronomy, Northwestern University, Evanston, IL 60208
[4]Center for Nanoscale Science and Technology, National Institute of Standards and Technology, Gaithersburg, MD 20899

## S1. EXPERIMENTAL DETAILS

Silicon microdisk cavities are fabricated in a silicon-on-insulator wafer with a 260 nm thick Si layer, 1 $\mu$m thick buried silicon dioxide layer, and specified device layer resistivity of 13.5 ohm-cm to 22.5 ohm-cm (p-type). Fabrication steps included electron-beam lithography of a 350 nm-thick positive-tone resist, an $SF_6/C_4F_8$ inductively-coupled plasma reactive ion etch through the silicon layer, a stabilized $H_2SO_4/H_2O_2$ etch to remove the remnant resist and other organic materials, and an HF wet etch to undercut the disks.

Devices were characterized (Fig. S1(a)) using a swept-wavelength external cavity tunable diode laser with a time-averaged linewidth less than 90 MHz and absolute stepped wavelength accuracy of 1 pm. Light is coupled into and out of the cavities using an optical fiber taper waveguide in a $N_2$-purged environment at atmospheric pressure and room temperature. Cavity transmission spectra were recorded using a InGaAs photoreceiver, while radio frequency (RF) spectra were recorded using a 0 MHz (DC) to 125 MHz InGaAs photoreceiver whose output was sent into an electronic spectrum analyzer.

Figure S1(b) shows a typical RF spectrum for a microdisk pumped with $P_{in} \approx 400$ $\mu$W at a fixed laser-cavity detuning, while Fig. S1(c),(d) compiles a series of such spectra as a function of laser-cavity detuning. Spectra such as these are analyzed to produce the period and amplitude data in Figs. 5 and 6 in the main text. Spectra are not shown for detunings where no oscillations occur (i.e., where the time domain signal is constant).

The discontinuity in Fig. S1(c) is a result of a transition between two resonant modes in the microdisk cavity. Data presented in Fig. 5 of the main text is restricted to a single mode consistent with our model.

## S2. PHYSICAL ORIGIN OF EQUATIONS

The section titled "Physical system and model" in the main text gives a concise explanation of the physical origin of Eqs. (1a)–(1c). Here, we expand on this by presenting a schematic of the model in Figure S2 and an explanation of all the model variables and parameters in Table S1. For a more extensive derivation, however, we refer the interested reader to the 2006 publication by Johnson, Borselli and Painter [1].

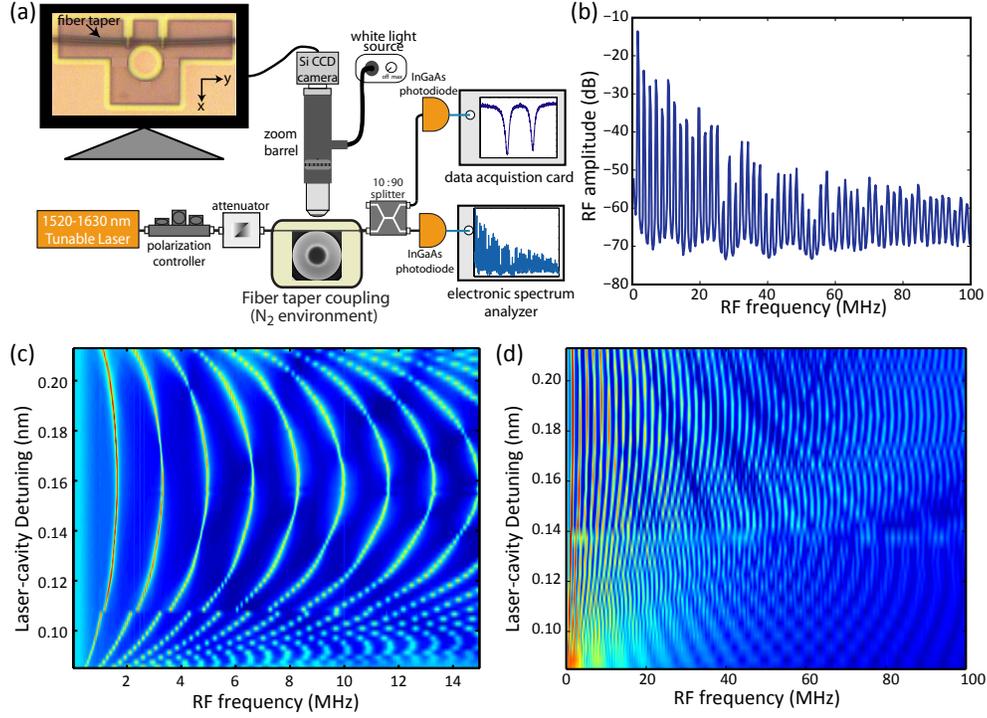

FIG. S1. (a) Experimental setup for measuring the silicon microdisk cavities. (b) RF spectrum of a microdisk for $P_{\text{in}} \approx 400\ \mu\text{W}$ and fixed laser-cavity detuning. (c)-(d) RF spectra as a function of laser-cavity detuning (laser frequency is detuned below cavity resonance frequency).

| Parameter | Meaning | Parameter | Meaning |
|---|---|---|---|
| $c$ | Speed of light | $V_{eff}$ | Volume of disk effectively occupied by resonant mode |
| $\hbar$ | Planck's constant / $2\pi$ | $V_{TPA}$ | Disk volume effectively available for two-photon absorption |
| $\lambda_0$ | Resonant wavelength of cavity mode | $V_{FCA}$ | Disk volume effectively available for free carrier absorption |
| $n_{Si}$ | Index of refraction for Si | $V_{disk}$ | Physical volume of microdisk |
| $n_g$ | Group index for the optical cavity mode | $Q$ | Intrinsic quality factor for disk at low optical powers ($Q = \omega_0/\gamma_0$) |
| $c_p$ | Heat capacity of Si | $\omega_0$ | Resonant angular frequency of mode ($\lambda_0 = 2\pi c/\omega_0$) |
| $\sigma_{Si}$ | Free carrier absorption cross section | $\gamma_0$ | Decay rate of EM field due to radiation and linear absorption |
| $\beta_{Si}$ | Two-photon absorption parameter | $\gamma_e$ | Decay rate of EM field due to coupling to the access waveguide ("extrinsic") |
| $\rho_{Si}$ | Density of Si | $\gamma_{lin}$ | Decay rate of the EM field due to linear optical absorption |
| $\frac{dn_{Si}}{dN}$ | Free-carrier effect on index of refraction | $\gamma_{fc}$ | Free carrier decay rate (inverse of free-carrier lifetime) |
| $\frac{dn_{Si}}{dT}$ | Temperature effect on index of refraction | $\gamma_{Th}$ | Thermal decay rate (inverse of thermal lifetime) |
| $\Gamma_{disk}$ | Fractional energy overlap with $\Delta T$ within the microdisk | $P_{in}$ | Power input (optical) |
| $\Gamma_{TPA}$ | Overlap factor for two-photon absorption | $\delta\omega_0$ | Detuning of input signal from resonance |
| $\Gamma_{FCA}$ | Overlap factor for free-carrier absorption | $\kappa$ | Coupling loss between fiber and disk |

TABLE S1. Definitions of variables and parameters from model.

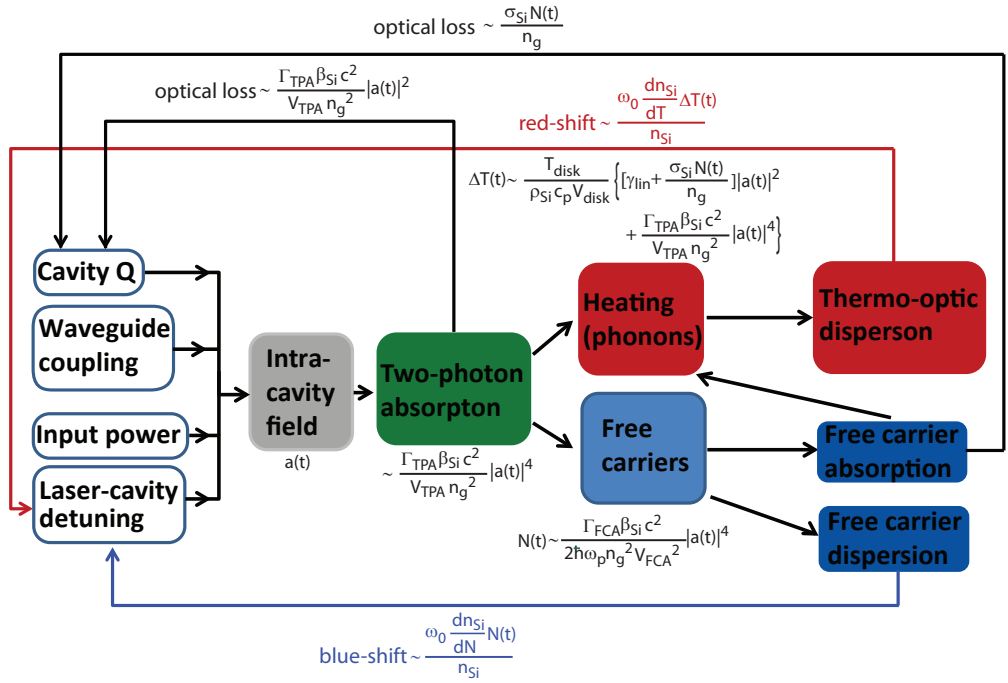

FIG. S2. Illustration of the different physical processes considered in the model, as represented by the equations of motion (Eqs. (1a)-(1c) in the main text). The intracavity optical field $a(t)$ is determined by the input optical power, waveguide coupling rate, laser-cavity detuning, and cavity $Q$ factor. A strong intracavity field leads to two-photon absorption, which reduces the cavity $Q$ while also generating heat and free carriers. The added heat produces thermo-optic dispersion and results in a red-shift of the cavity mode with respect to the laser, while the generated free carriers lead to absorption and dispersion. Free-carrier absorption reduces the cavity $Q$ while also potentially leading to additional heating and thermo-optic dispersion, while free-carrier dispersion results in a blue shift of the cavity with respect to the laser.

## S3. DIMENSIONAL CONSTANTS:

Table S2 lists the dimensional constants used in the model. Values given are those used in calculation unless otherwise specified.

| Parameter | Value | Source | Parameter | Value | Source |
|---|---|---|---|---|---|
| $c$ | $2.998 \times 10^8$ m/s | physical const. | $V_{eff}$ | $50(\frac{\lambda_0}{n_{Si}})^3 = 3.99 \times 10^{-18}$ m$^3$ | FEM [1] * |
| $\hbar$ | $1.05 \times 10^{-34}$ J s | physical const. | $V_{TPA}$ | $2V_{eff} = 7.97 \times 10^{-18}$ m$^3$ | FEM [1] * |
| $\lambda_0$ | $1.5 \times 10^{-6}$ m | low-power meas. * | $V_{FCA}$ | $2V_{eff} = 7.97 \times 10^{-18}$ m$^3$ | FEM [1] * |
| $n_{Si}$ | 3.485 | material const. | $V_{disk}$ | $10V_{eff} = 3.99 \times 10^{-17}$ m$^3$ | FEM [1] * |
| $n_g$ | 3.485 | material const. | $Q$ | $3 \times 10^5$ | low-power meas. * |
| $c_p$ | 700 J/(kg K) | material const. | $\omega_0$ | $\frac{2\pi c}{\lambda_0} = 1.26 \times 10^{15}$ Hz | low-power meas. * |
| $\sigma_{Si}$ | $10^{-21}$ m$^2$ | material const. | $\gamma_0$ | $\frac{2\pi c}{\lambda_0 Q} = 4.19 \times 10^9$ Hz | low-power meas. * |
| $\beta_{Si}$ | $8.4 \times 10^{-12}$ m/W | material const. | $\gamma_e$ | $\frac{2\pi c}{\lambda_0 Q} = 4.19 \times 10^9$ Hz | critical coupling * |
| $\rho_{Si}$ | 2330 kg/m$^3$ | material const. | $\gamma_{lin}$ | $\frac{2\pi c}{\lambda_0 Q} = 4.19 \times 10^9$ Hz | no radiation loss * |
| $\frac{dn_{Si}}{dN}$ | $-1.73 \times 10^{-27}$m$^3$ | material const. | $\gamma_{fc}$ | $10^8$ Hz | [1] |
| $\frac{dn_{Si}}{dT}$ | $1.86 \times 10^{-4}$ K$^{-1}$ | material const. | $\gamma_{Th}$ | $2 \times 10^5$ Hz | [1] * |
| $\Gamma_{disk}$ | 1 | FEM [1] | $P_{in}$ | in range 30 $\mu$W $-$ 3160 $\mu$W | measured |
| $\Gamma_{TPA}$ | 1 | FEM [1] | $\delta\omega_0$ | in range 0 Hz $-$ 3.3 $\times 10^{11}$ Hz | measured |
| $\Gamma_{FCA}$ | 1 | FEM [1] | $\kappa$ | $\sqrt{\gamma_e} = 6.47 \times 10^5$ $\sqrt{\text{Hz}}$ | critical coupling * |

TABLE S2. Constants used for simulation and analysis. Values of parameters (*) used for comparison to laboratory data in Fig. 5 of the main text are given in figure caption.

## S4. NONDIMENSIONAL CONSTANTS:

Consider the nondimensional equations

$$\frac{dU}{d\tau} = -A_1 U - A_2 S^2 U(U^2 + V^2) - A_3 \eta U + A_4 \eta V + A_5 x V - A_6 \theta V \ , \tag{S1a}$$

$$\frac{dV}{d\tau} = -A_1 V - A_2 S^2 V(U^2 + V^2) - A_3 \eta V - A_4 \eta U - A_5 x U + A_6 \theta U - A_7 \ , \tag{S1b}$$

$$\frac{d\eta}{d\tau} = -A_8 \eta + A_9 S^4 (U^2 + V^2)^2 \ , \tag{S1c}$$

$$\frac{d\theta}{d\tau} = -A_{10} \theta + A_{11} S^2 (U^2 + V^2) + A_{12} S^4 (U^2 + V^2)^2 + A_{13} S^2 \eta (U^2 + V^2) \ , \tag{S1d}$$

resulting from a nondimensionalization of the form

$$\tau = c_1 t,$$

$$U(\tau) = \frac{1}{c_2} \mathrm{Re}(a(t)),$$

$$V(\tau) = \frac{1}{c_2} \mathrm{Im}(a(t)),$$

$$\eta(\tau) = \frac{1}{c_3} N(t),$$

$$\theta(\tau) = \frac{1}{c_4} \Delta T(t),$$

with

$$c_1 = \frac{\gamma_0}{\sqrt{Q}},$$

$$c_2 = \frac{6 Q^{1/4} \sqrt{P_{in}}}{\sqrt{\omega_0}},$$

$$c_3 = \frac{Q}{V_{eff}},$$

$$c_4 = \frac{\gamma_0^2 \sigma_{Si} Q}{c_p}.$$

The system is non-dimensionalized with a characteristic time, length, mass, and temperature scale. Here we have taken

$$[\text{time}] = \gamma_0^{-1}$$

$$[\text{length}] = V_{eff}^{1/3}$$

$$[\text{mass}] = \frac{P_{in}}{\gamma_0^3 V_{eff}^{2/3}}$$

$$[\text{temp.}] = \frac{\gamma_0^2 \sigma_{Si}}{c_p},$$

The quality factor $Q$ was used to scale the dynamic variables to ranges of $\mathcal{O}(1)$. $A_1$ through $A_{13}$ are positive real constants, given as follows:

| Control Parameter | Expression | Range |
|---|---|---|
| $x$ | $\frac{\delta\omega_0}{\gamma_0}$ | 0-500, typical value 100 |
| $S$ | $\frac{sQ\sqrt{\beta_{Si}\omega_0}}{\sqrt{c}}$ | 0-100, typical value 56 |
| Nondimensional Coefficient | Expression | Value |
| $A_1$ | $\frac{\gamma_0+\gamma_c}{2}\frac{\sqrt{Q}}{\gamma_0}$ | 547.72 |
| $A_2 S^2$ | $\left(\frac{18\Gamma_{TPA}c^3}{V_{TPA}n_g^2\gamma_0\omega_0^2 Q}\right)S^2$ | $0.0025289\ S^2$, typical value 7.9307 |
| $A_3$ | $\frac{\sigma_{Si}cQ^{3/2}}{2\gamma_0 n_g V_{eff}}$ | 423.49 |
| $A_4$ | $\frac{-\omega_0\frac{dn_{Si}}{dN}Q^{3/2}}{n_{Si}\gamma_0 V_{eff}}$ | 6137.8 |
| $A_5 x$ | $\left(\frac{\gamma_c\sqrt{Q}}{\gamma_0}\right)x$ | $547.72\ x$, typical value $5.4772\times10^4$ |
| $A_6$ | $\frac{\omega_0\frac{dn_{Si}}{dT}\gamma_0\sigma_{Si}Q^{3/2}}{n_{Si}c_p}$ | $6.5858\times10^4$ |
| $A_7$ | $\frac{\kappa Q^{1/4}\sqrt{\omega_0}}{6\gamma_0}$ | 2136.4 |
| $A_8$ | $\frac{\gamma_{fc}\sqrt{Q}}{\gamma_0}$ | 13.085 |
| $A_9 S^4$ | $\left(\frac{648\Gamma_{FCA}V_{eff}c^4}{\beta_{Si}\omega_0^4 Q^{7/2}\hbar\omega_0\gamma_0 n_g^2 V_{FCA}^2}\right)S^4$ | $1.5849\times10^{-7}\ S^4$, typical value 1.5589 |
| $A_{10}$ | $\frac{\gamma_{Th}\sqrt{Q}}{\gamma_0}$ | 0.026152 |
| $A_{11} S^2$ | $\left(\frac{36T_{disk}\gamma_{lin}c}{\rho_{Si}V_{disk}\beta_{Si}\omega_0^2 Q^2\gamma_0^3\sigma_{Si}}\right)S^2$ | $5.5614\times10^{-6}\ S^2$, typical value 0.017440 |
| $A_{12} S^4$ | $\left(\frac{1296T_{disk}\Gamma_{TPA}c^4}{\beta_{Si}\rho_{Si}V_{disk}V_{TPA}n_g^2\omega_0^4 Q^{7/2}\gamma_0^3\sigma_{Si}}\right)S^4$ | $5.1356\times10^{-11}\ S^4$, typical value $5.0506\times10^{-4}$ |
| $A_{13} S^2$ | $\left(\frac{36T_{disk}c^2}{\rho_{Si}V_{disk}n_g\beta_{Si}\omega_0^2 QV_{eff}\gamma_0^3}\right)S^2$ | $8.6000\times10^{-6}\ S^2$, typical value 0.026970 |

TABLE S3. Non-dimensional parameters and their typical numerical values. $x$ and $S$ are control parameters while $A_{1-13}$ are fixed parameters.

## S5. BIRTH OF LIMIT CYCLE

Consider a limit cycle of the state $\vec{\psi}(\tau)=(U,\ V,\ \eta,\ \theta)$ parameterized by $(x,S)$. Call the region of parameter space where this limit cycle exists $\Sigma$. Define the limit cycle $\vec{\psi}(t)=\vec{\psi}(t+T)$ as $L_{x,S}$, with period $T$.

The limit cycle is "born" in parameter space on the boundary $\partial\Sigma$, defined by a Hopf-condition with zero amplitude and non-zero period $T$. The Hopf-condition is the requirement that a pair of complex conjugate eigenvalues cross the imaginary axis. The limit cycle is either born stable (supercritical) or unstable (subcritical). We have identified numerically the point at which stability changes, which is visible in Fig. 3 (asterisk near $P_{in}=125\ \mu$W, $\delta\omega_0/\gamma_0=60$).

The limit cycle is born as a sinusoidal wave for each variable in time. To lowest order state variable $\vec{\psi}_i\propto\sin(\frac{2\pi t}{T}+\alpha_i)$, where $\alpha_i$ is a phase lag parameter to be determined. In system (S1), define the difference in the parameters $x$ and $S$ from the Hopf bifurcation to be

$$x-x_{Hopf}=\nu_1\epsilon^2$$
$$S-S_{Hopf}=\nu_2\epsilon^2,$$

where $\epsilon\ll 1$ and $\nu_1$ and $\nu_2$ are $\mathcal{O}(1)$. Assume the limit cycle can be expressed as a power series in $\epsilon$. As $\epsilon\to 0$, we expect the expansion to approach the true limit cycle asymptotically. The following series assumes that each variable can be expanded as a sum of a homogeneous part and a sinusoidal part, and was found to conveniently solve the system of equations in the limit of small $\epsilon$:

$$U(\tau|\epsilon) = U_H + U_{osc} = \sum_{n=0}^{\infty} \epsilon^{2n} U_n + \sum_{n=1}^{\infty} a_n \epsilon^n \cos^n (\omega\tau + \alpha_n)$$

$$\text{where} \quad \omega = \omega(\epsilon) = \sum_{k=0}^{\infty} \omega_k \epsilon^{2k}, \quad \alpha_n = \alpha_n(\epsilon) = \sum_{k=0}^{\infty} \alpha_{n,k} \epsilon^{2k}$$

$$V(\tau|\epsilon) = V_H + V_{osc} = \sum_{n=0}^{\infty} \epsilon^{2n} V_n + \sum_{n=1}^{\infty} b_n \epsilon^n \cos^n (\omega\tau + \beta_n)$$

$$\text{where} \quad \omega = \omega(\epsilon) = \sum_{k=0}^{\infty} \omega_k \epsilon^{2k}, \quad \beta_n = \beta_n(\epsilon) = \sum_{k=0}^{\infty} \beta_{n,k} \epsilon^{2k}$$

$$\eta(\tau|\epsilon) = \eta_H + \eta_{osc} = \sum_{n=0}^{\infty} \epsilon^{2n} \eta_n + \sum_{n=1}^{\infty} c_n \epsilon^n \cos^n (\omega\tau + \gamma_n)$$

$$\text{where} \quad \omega = \omega(\epsilon) = \sum_{k=0}^{\infty} \omega_k \epsilon^{2k}, \quad \gamma_n = \gamma_n(\epsilon) = \sum_{k=0}^{\infty} \gamma_{n,k} \epsilon^{2k}$$

$$\theta(\tau|\epsilon) = \theta_H + \theta_{osc} = \sum_{n=0}^{\infty} \epsilon^{2n} \theta_n + \sum_{n=1}^{\infty} d_n \epsilon^n \cos^n (\omega\tau + \delta_n)$$

$$\text{where} \quad \omega = \omega(\epsilon) = \sum_{k=0}^{\infty} \omega_k \epsilon^{2k}, \quad \delta_n = \delta_n(\epsilon) = \sum_{k=0}^{\infty} \delta_{n,k} \epsilon^{2k}.$$

To 3rd order in $\epsilon$, for example, the $U(\tau)$ expansion becomes,

$$U(\tau|\epsilon) = U_0 + \epsilon^2 U_1 + (a_{1,0} + \epsilon^2 a_{1,1})\epsilon \cos\left((\omega_0 + \epsilon^2\omega_1)\tau + \alpha_{1,0} + \epsilon^2\alpha_{1,1}\right)$$
$$+ a_{2,0}\epsilon^2 \cos\left((\omega_0 + \epsilon^2\omega_1)\tau + \alpha_{2,0}\right)^2 + a_{3,0}\epsilon^3 \cos\left((\omega_0 + \epsilon^2\omega_1)\tau + \alpha_{3,0}\right)^3.$$

Solutions for the unknown constants, specifically the frequency $\omega$, amplitudes $a$, $b$, $c$, $d$, and phase shift $\alpha$, $\beta$, $\gamma$, $\delta$, are found through substitution of the form into the equations of motion (S1). The only free parameter is $\epsilon$. Thus, the form of the limit cycle can be accurately approximated close to the Hopf condition. Figures S3, S4, and S5 show typical examples of the approximation versus the numerics, where $R = \sqrt{U^2 + V^2}$ is the nondimensional magnitude of the field. The base Hopf point in this calculation was $x = 60$ (0.3 nm), $S \approx 17.37$ (95 μW), and the perturbation in parameter space was taken to be in $S$ with fixed $x$ ($\nu_1 = 0$, $\nu_2 = 1$). (Note: Here we have carried out the expansion to 5th order, and determined the period of oscillation to third order and the solution to 2nd order.)

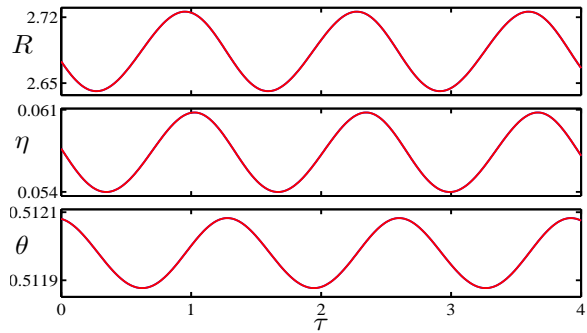

FIG. S3. Numerical solution (blue) and asymptotic expansion (red) of limit cycle near Hopf location. Here $\epsilon^2 = 0.01$, $S = S_{hopf} + \epsilon^2 \approx 17.38$ (95 μW), $x = 60$ (0.3 nm). Note that red curve obscures blue curve.

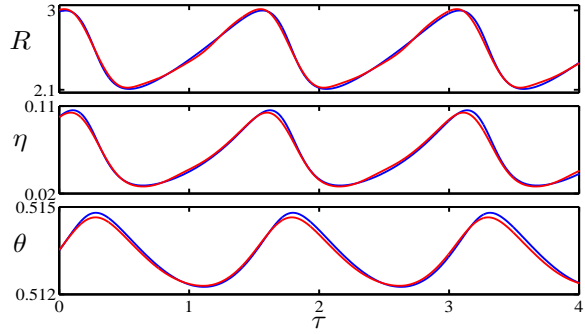

FIG. S4. Numerical solution (blue) and asymptotic expansion (red) of limit cycle near Hopf location. Here $\epsilon^2 = 0.9$, $S = S_{hopf} + \epsilon^2 \approx 18.27$ (105 $\mu$W), $x = 60$ (0.3 nm).

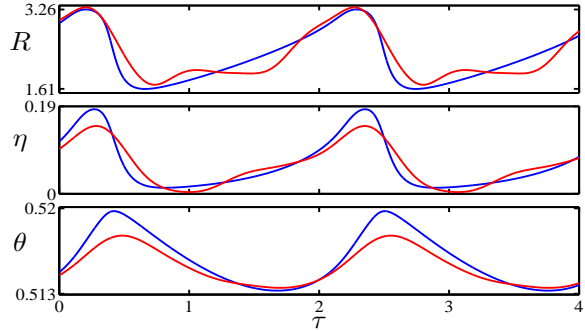

FIG. S5. Numerical solution (blue) and asymptotic expansion (red) of limit cycle near Hopf location. Here $\epsilon^2 = 2$, $S = S_{hopf} + \epsilon^2 \approx 19.37$ (118 $\mu$W), $x = 60$ (0.3 nm). Note that while the shape of the asymptotic limit cycle diverges from the numerics, the period remains relatively accurate.

Figure S6 compares the local approximation for the period of oscillation with the period from numerical integration.

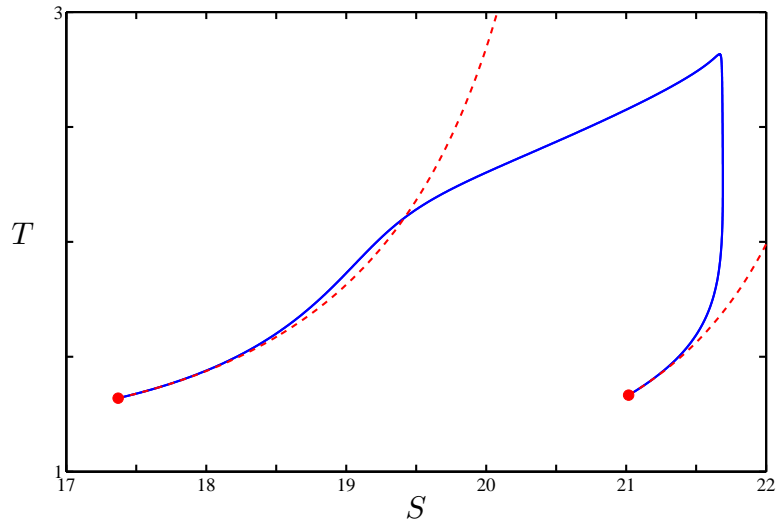

FIG. S6. Period of limit cycle $T$ vs power $S$, from numerical solution (blue) and asymptotic expansions (red dashed). Two separate expansions from the Hopf locations (red circles) are shown: The limit cycle is born stable (supercritical) on the left, and unstable (subcritical) on the right. Here $x = 60$ (0.3 nm).

The asymptotic expansion yields locally accurate approximations for the limit cycle's shape and period. This fit becomes less accurate as the nonlinearity of the oscillations increases.

A limitation of this approach is that it cannot predict the occurrence of the homoclinic bifurcation in parameter space. We found that the multiple time scale analysis yields more useful insight as the strength of the input signal is increased.

## S6. SURVEY OF PERIOD'S DEPENDENCY UPON PARAMETERS

Table S4 graphically depicts typical dependence of the full microdisk oscillation period on various physical parameters.

The time scale of the "spike" (red section in Fig. 2) increases linearly with $V_{eff}$, decays approximately with input power like $P_{in}^{-1}$, decays approximately with quality factor like $Q^{-1}$, and is unaffected by changes in detuning and $\gamma_{Th}$. These scalings are based upon analytic approximations of the 1D model, and are confirmed by numerical integration of the 2D model.

## S7. MULTIPLE TIME SCALE ANALYSIS

### Time Scale Separation

The separation of time scales can be formalized as the ratio of decay rates $\gamma$ of the field ($U$, $V$), free carriers ($\eta$) and temperature ($\theta$). Scaling system (S1) by the nondimensional parameter $A_1^{-1}$, and scaling time as $\tau_1 = A_1 \tau = \gamma_0 t$ gives,

$$\frac{dU}{d\tau_1} = -U - \frac{A_2 S^2}{A_1} U(U^2 + V^2) - \frac{A_3}{A_1}\eta U + \frac{A_4}{A_1}\eta V + \frac{A_5}{A_1}xV - \frac{A_6}{A_1}\theta V \ , \tag{S5a}$$

$$\frac{dV}{d\tau_1} = -V - \frac{A_2 S^2}{A_1} V(U^2 + V^2) - \frac{A_3}{A_1}\eta V - \frac{A_4}{A_1}\eta U - \frac{A_5}{A_1}xU + \frac{A_6}{A_1}\theta U - \frac{A_7}{A_1} \ , \tag{S5b}$$

$$\frac{d\eta}{d\tau_1} = -\frac{\gamma_{fc}}{\gamma_0}\left(\eta - \frac{A_9 S^4}{A_8}(U^2 + V^2)^2\right) \ , \tag{S5c}$$

$$\frac{d\theta}{d\tau_1} = -\frac{\gamma_{Th}}{\gamma_0}\left(\theta - \frac{A_{11}S^2}{A_{10}}(U^2 + V^2) - \frac{A_{12}S^4}{A_{10}}(U^2 + V^2)^2 - \frac{A_{13}S^2}{A_{10}}\eta(U^2 + V^2)\right) \ . \tag{S5d}$$

If all the coefficients on the right hand side in one differential equation in the system are orders of magnitude larger than all coefficients of another differential equation in the system and all variables are of similar order, the first dynamic variable is said to evolve on a faster time scale. Equations (S5a) and (S5b) evolve on a faster time scale than the Eqs. (S5c) and (S5d) if the following conditions are met:

$$\frac{\gamma_{fc}}{\gamma_0} \ll 1, \tag{S6}$$

$$\frac{\gamma_{Th}}{\gamma_0} \ll 1, \tag{S7}$$

$$\frac{A_{3,4,5,6,7}}{A_1} \gtrsim 1 \tag{S8}$$

$$\frac{A_2 S^2}{A_1} \gtrsim 1 \tag{S9}$$

$$\frac{A_9 S^4}{A_8} \lesssim 1 \tag{S10}$$

$$\frac{A_{11}S^2}{A_{10}}, \frac{A_{12}S^4}{A_{10}}, \frac{A_{13}S^2}{A_{10}} \lesssim 1. \tag{S11}$$

For the devices considered in this work, conditions (S6), (S7), (S8), (S10), and (S11) are met, while condition (S9) is not (see Table S3). Nevertheless, we can treat Eqs. (S5a) and (S5b) as evolving on separate time scales from Eqs. (S5c) and (S5d) (we will address this in the next subsection).

We also observe that Eq. (S5c) evolves on a faster time scale than Eq. (S5d). Dividing Eqs. (S5c) and (S5d) by

| Parameter | General Behavior of Period | Figure |
|:---:|:---:|:---:|
| $\delta\omega_0/\gamma_0$ | Non-monotonic | 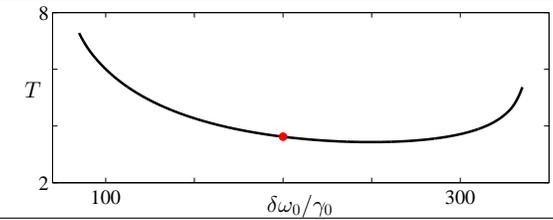 |
| $S$ | Increasing | 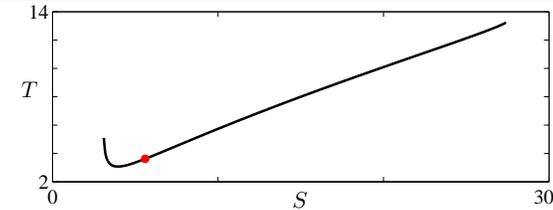 |
| $Q$ | Decreasing | 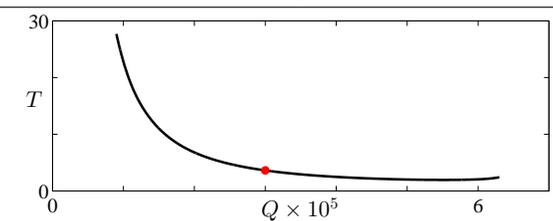 |
| $V_{eff}$ | Non-monotonic | 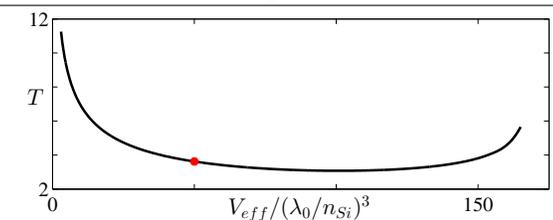 |
| $\gamma_e/\gamma_0$ | Non-monotonic | 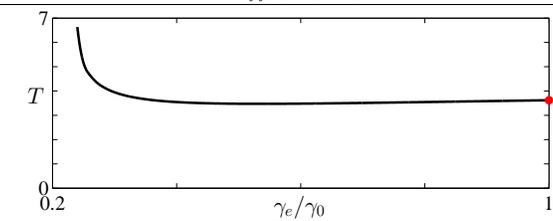 |
| $\gamma_{fc}$ | Decreasing | 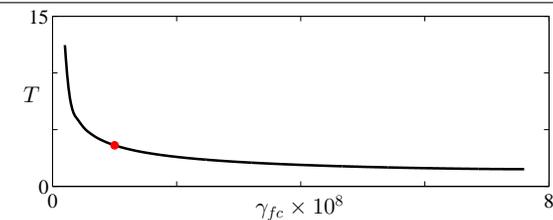 |
| $\gamma_{Th}$ | Non-monotonic | 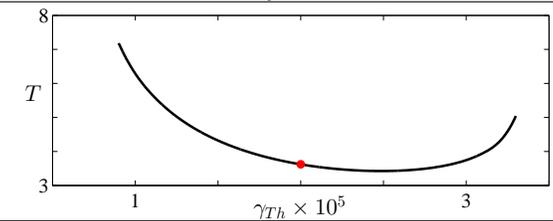 |

TABLE S4. Numerical survey of dependency of the nondimensional period of the limit cycle in parameter space. The red circle in each plot corresponds to values of the parameters given in Table S2, with $\delta\omega_0/\gamma_0 = 200$ and $P_{in} = 1$ mW ($S = 56$).

$\gamma_{fc}/\gamma_0$, and rescaling time as $\tau_2 = (\gamma_{fc}/\gamma_0)\tau_1 = \gamma_{fc}t$ yields the following system:

$$\frac{dU}{d\tau_1} = -U - \frac{A_2 S^2}{A_1} U(U^2 + V^2) - \frac{A_3}{A_1}\eta U + \frac{A_4}{A_1}\eta V + \frac{A_5}{A_1}xV - \frac{A_6}{A_1}\theta V \ , \tag{S12a}$$

$$\frac{dV}{d\tau_1} = -V - \frac{A_2 S^2}{A_1} V(U^2 + V^2) - \frac{A_3}{A_1}\eta V - \frac{A_4}{A_1}\eta U - \frac{A_5}{A_1}xU + \frac{A_6}{A_1}\theta U - \frac{A_7}{A_1} \ , \tag{S12b}$$

$$\frac{d\eta}{d\tau_2} = -\eta + \frac{A_9 S^4}{A_8}(U^2 + V^2)^2 \ , \tag{S12c}$$

$$\frac{d\theta}{d\tau_2} = -\frac{\gamma_{Th}}{\gamma_{fc}}\left(\theta - \frac{A_{11} S^2}{A_{10}}(U^2 + V^2) - \frac{A_{12} S^4}{A_{10}}(U^2 + V^2)^2 - \frac{A_{13} S^2}{A_{10}}\eta(U^2 + V^2)\right) \ . \tag{S12d}$$

The time scales of Eqs. (S12c) and (S12d) can be formally separated if the following conditions are met:

$$\frac{\gamma_{Th}}{\gamma_{fc}} \ll 1, \tag{S13}$$

$$\frac{A_8}{A_9 S^4} \lesssim 1 \tag{S14}$$

$$\frac{A_{11} S^2}{A_{10}}, \frac{A_{12} S^4}{A_{10}}, \frac{A_{13} S^2}{A_{10}} \lesssim 1 \qquad \text{(c.f. Eq. (S11))} \tag{S15}$$

For the devices considered in this work, conditions (S13), (S14), and (S15) are met. There is, of course, some error in this reduction, apparent in Fig. 4 of the main text: some change in $\theta$ is visible when $\eta$ is not on its nullcline, $\frac{d\eta}{d\tau} = 0$.

The full separation of time scales yields the system

$$\frac{dU}{d\tau_1} = -U - \frac{A_2 S^2}{A_1} U(U^2 + V^2) - \frac{A_3}{A_1}\eta U + \frac{A_4}{A_1}\eta V + \frac{A_5}{A_1}xV - \frac{A_6}{A_1}\theta V \ , \tag{S16a}$$

$$\frac{dV}{d\tau_1} = -V - \frac{A_2 S^2}{A_1} V(U^2 + V^2) - \frac{A_3}{A_1}\eta V - \frac{A_4}{A_1}\eta U - \frac{A_5}{A_1}xU + \frac{A_6}{A_1}\theta U - \frac{A_7}{A_1} \ , \tag{S16b}$$

$$\frac{d\eta}{d\tau_2} = -\eta + \frac{A_9 S^4}{A_8}(U^2 + V^2)^2 \ , \tag{S16c}$$

$$\frac{d\theta}{d\tau_3} = -\theta + \frac{A_{11} S^2}{A_{10}}(U^2 + V^2) + \frac{A_{12} S^4}{A_{10}}(U^2 + V^2)^2 + \frac{A_{13} S^2}{A_{10}}\eta(U^2 + V^2) \ , \tag{S16d}$$

where $\tau_1 = \gamma_0 t$, $\tau_2 = \gamma_{fc}t$, and $\tau_3 = \gamma_{Th}t$, with $\tau_1 \gg \tau_2 \gg \tau_3$.

### Limitations of Multiple Times Scales

*Condition S9*—Numerically, removing the $A_2 S^2$ term in Eqs. (S1a) and (S1b) has negligible effect on the system's fixed points and dynamic behavior (the accuracy of the 2D reduction in representing the 4D model is further confirmation of this assertion). Thus, we will remove the terms in Eqs. (S1a) and (S1b) with coefficient $A_2 S^2$ in our multiple time scale analysis. This is reasonable, given that the magnitude of the $A_2 S^2$ is much smaller than other terms in Eqs. (S1a) and (S1b). (We have quantified the error introduced by removing this term and found that it is $\mathcal{O}(\frac{A_2 S^2}{A_1})$.) At a power higher than $S = 150$, ($P_{in} = 7$ mW), the term $A_2 S^2$ is within an order of magnitude of $A_1$, and thus we would expect this nonlinearity to become more important. In addition, due to its nonlinearity, we expect this term to affect the spike during the limit cycle, and our model excluding this term to be least accurate at the spiking event.

In the regime of low power (low $S$) condition (S14) breaks down. Using the standard parameter values given in Table S3, we find $S \gtrsim 95$, or that this condition is satisfied when $P_{in} > 3$ mW. This lower bound for $P_{in}$ is actually more strict than necessary for a satisfactory separation of times scales: comparing the magnitude of $A_9 S^4$ to the largest term in Eq. (S1d) gives a more generous condition. Assuming an order of magnitude difference between the terms suggests the time scale separation begins to break down at $S \approx 35$, or powers lower than approximately 380 $\mu$W.

In the regime of high power, (high $S$), conditions (S10) and (S11) breaks down. The strictest of these four relations for our parameter values is (S10). The term $A_9 S^4$ is an order of magnitude greater than $A_8$ when $S > 170$, ($P_{in} = 9$mW).

## 1D Model

After reducing the 4D system to a 2D system, as described in the main text, we can further reduce to a 1D model, as presented in Eqs. (4) and (5):

$$\theta = \frac{A_5 x + A_4 \eta}{A_6} \pm \frac{\sqrt{-A_3^2 \eta^4 - 2A_1 A_3 \eta^3 - A_1^2 \eta^2 + \sqrt{\frac{A_7^4 A_9 S^4}{A_8}} \eta^{3/2}}}{A_6 \eta}, \tag{S17}$$

$$\dot{\theta} = -\frac{A_{10} A_5 x}{A_6} + \frac{\sqrt{A_8} A_{11}}{\sqrt{A_9}} \sqrt{\eta} + \left( \frac{A_8 A_{12}}{A_9} - \frac{A_{10} A_4}{A_6} \right) \eta \tag{S18}$$

$$+ \frac{\sqrt{A_8} A_{13}}{\sqrt{A_9}} \eta^{3/2} \pm \frac{A_{10} \sqrt{-A_3^2 \eta^4 - 2A_1 A_3 \eta^3 - A_1^2 \eta^2 + \sqrt{\frac{A_7^4 A_9 S^4}{A_8}} \eta^{3/2}}}{A_6 \eta}.$$

This is a parameterized 1D system with two branches (represented by the $\pm$) that approximate the two slow sections of the limit cycle. The boundaries of the one dimensional limit cycle are the transition points between these two solution branches, which are defined by the condition

$$\frac{d\theta}{d\eta} = 0. \tag{S19}$$

Converting the condition to a polynomial expresses the critical $\eta$ values as roots to a tenth order polynomial. Note from Eq. (S17) that the expression is independent of $x$. This allows us to numerically solve the expression while retaining dependence upon the control parameter $x$. This "critical $\eta$" condition has two real, positive solutions, $\eta_1^*$ and $\eta_3^*$. We can also find the "collection" points on each of the two branches that the solution jumps to, $\eta_2^*$ and $\eta_4^*$. Plugging the solutions for $\eta$ into (S17) gives the maximum and minimum $\theta$ values of the limit cycle, and into Eq. (S18) the corresponding $\dot{\theta}$ values. The $\eta$ solutions have no dependence upon detuning $x$, while $\theta$ and $\dot{\theta}$ have linear dependence upon detuning:

$$\theta_{max}^* = \theta_1^* = \frac{A_5 x}{A_6} + g_1(S), \tag{S20a}$$

$$\theta_{min}^* = \theta_3^* = \frac{A_5 x}{A_6} + g_2(S), \tag{S20b}$$

$$\dot{\theta}_i^* = \frac{-A_{10} A_5 x}{A_6} + h_i(S), \tag{S20c}$$

where $g_1$, $g_2$, and $h_i$ are implicit functions of $S$ defined as roots of a polynomial. Figure 4 in the main text labels these points for the 1D limit cycle.

Figure S7 shows the dependency of the shape of the 1D limit cycle upon the control parameters. Shifts in detuning translate the limit cycle without changing its shape. Changes in driving power both translate and adjust the shape of the limit cycle.

## Analytic Approximation to 1D Model

The functions $g_{1,2}$ and $h_i$ from Eq. (S20) implicitly depend upon all the problem parameters $A_1$ through $A_{13}$. We can approximate that dependence using Taylor series near the critical $\eta$ condition. According to our nondimensionalization, we expect that $\eta_1^*$ will be $\mathcal{O}(1)$, while the $\eta_3^*$ will be very small. Under these assumptions, we find

$$\eta_1^* \approx \frac{A_3 - 3A_1}{A_1 + 5A_3} + \frac{4A_7 A_9^{1/4} S}{A_8^{1/4}(A_1 + 5A_3)},$$

$$\eta_3^* \approx \left( \frac{A_7 A_9^{1/4} S}{4 A_8^{1/4} A_4} \right)^{4/5}.$$

Using these two values we can find expressions for all four locations on the limit cycle in terms of any desired parameter. For example, the onset of oscillations (Hopf bifurcation) occurs when $\dot{\theta}_3^* = 0$ for increasing detuning.

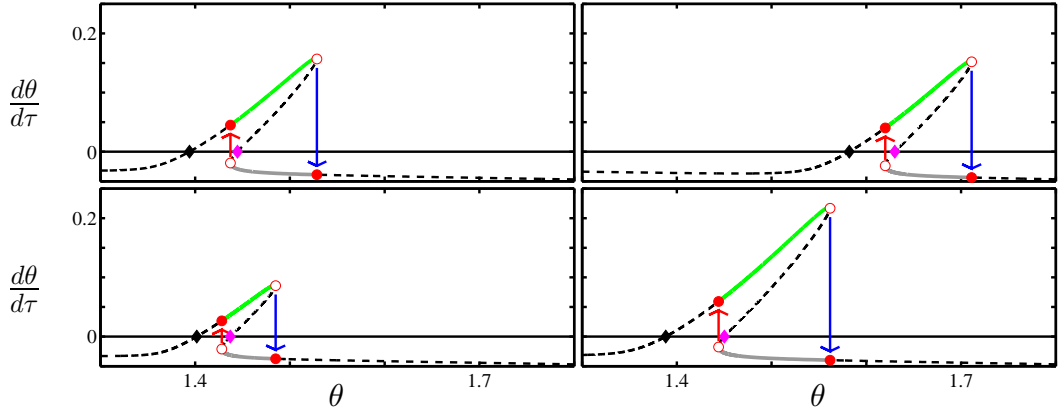

FIG. S7. Change of 1D limit cycle with respect to detuning and power. Shown are nullclines of $\eta$ (dashed), unstable fixed points (filled diamonds), and points of interest in the 1D reduction (filled and open circles). Arrows indicate instantaneous jumps (from time scale separation). Parameter values: ($S = 56$, $x = 168$), ($S = 56$, $x = 190$), ($S = 42$, $x = 168$), ($S = 67$, $x = 168$). Changes in detuning shift the limit cycle while changes in power elongate or collapse limit cycle branches.

Plugging the approximate expression for $\eta_3^*$ into Eq. (S18) yields a general expression of the Hopf condition for all control parameters in the problem. Similarly, the onset of oscillations (Hopf bifurcation) for decreasing detuning occurs when $\dot{\theta}_1^* = 0$. These approximations for the Hopf location are compared with the actual Hopf location from the 4D system and the 1D system in Fig. S8.

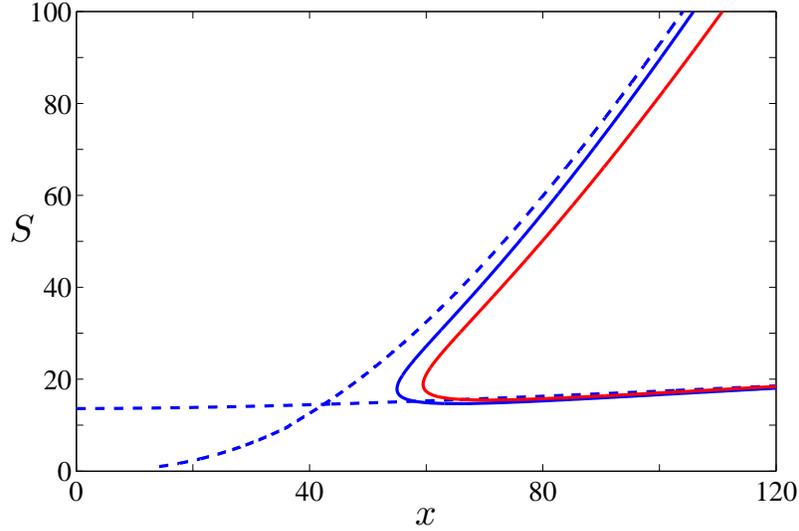

FIG. S8. Onset of oscillations with respect to detuning from resonance. The Hopf bifurcation of the four dimensional system (red) is compared with the onset of oscillations for the 1D system (blue) and the analytic approximation for the 1D system (dashed blue). Power ranges from 0 mW to 3.16 mW, Detuning ranges from 0 nm to 0.5 nm. Note that the homoclinic bifurcation is not shown.

* dmabrams@northwestern.edu
† AlexanderSlawik2015@u.northwestern.edu
‡ kartik.srinivasan@nist.gov



\end{document}